\documentclass{article}
\usepackage{spconf,amsmath,graphicx}
\usepackage{xcolor}
\usepackage{enumitem}
\usepackage{csvsimple}
\setlist{nosep, leftmargin=14pt}
\usepackage{booktabs}
\usepackage{hyperref,cleveref}
\hypersetup{colorlinks=true,linkcolor=[rgb]{0.1,0.3,0.9},urlcolor=[rgb]{0.2,0.2,0.2},citecolor=[rgb]{0.1,0.3,0.9}}

\makeatletter\newcommand*\mysizea{\@setfontsize\mysizea{6.25}{9.5}}\makeatother

\title{Deep Learning-Based Assessment of Cerebral Microbleeds in COVID-19}

\name{\begin{tabular}{@{}c@{}}
Neus Rodeja Ferrer $^{\star}$ \hspace{0.25em}
Malini Vendela Sagar $^{\dagger}$ \hspace{0.25em}
Kiril Vadimovic Klein $^{\star}$ \\
Christina Kruuse $^{\dagger}$ \hspace{0.25em}
Mads Nielsen $^{\star}$ \hspace{0.25em}
Mostafa Mehdipour Ghazi $^{\star}$
\end{tabular}}

\address{$^{\star}$ Department of Computer Science, University of Copenhagen \\ $^{\dagger}$ Department of Neurology, Copenhagen University Hospital}

\begin{document}

\maketitle

\begin{abstract}
Cerebral Microbleeds (CMBs), typically captured as hypointensities from susceptibility-weighted imaging (SWI), are particularly important for the study of dementia, cerebrovascular disease, and normal aging. Recent studies on COVID-19 have shown an increase in CMBs of coronavirus cases. Automatic detection of CMBs is challenging due to the small size and amount of CMBs making the classes highly imbalanced, lack of publicly available annotated data, and similarity with CMB mimics such as calcifications, irons, and veins. Hence, the existing deep learning methods are mostly trained on very limited research data and fail to generalize to unseen data with high variability and cannot be used in clinical setups. To this end, we propose an efficient 3D deep learning framework that is actively trained on multi-domain data. Two public datasets assigned for normal aging, stroke, and Alzheimer’s disease analysis as well as an in-house dataset for COVID-19 assessment are used to train and evaluate the models. The obtained results show that the proposed method is robust to low-resolution images and achieves 78\% recall and 80\% precision on the entire test set with an average false positive of 1.6 per scan.
\end{abstract}


% To start a new column (but not a new page) and help balance the last-page column length use: \vfill \pagebreak

\section{Introduction}
\label{sec:intro}
Cerebral Microbleeds (CMBs) are small foci of chronic blood products in normal brain tissue leaked from cerebral small vessels \cite{Greenberg2009}. They are recognized as neurological findings in patients with cerebrovascular disease \cite{Charidimou2012}, dementia \cite{Qiu2010}, normal aging \cite{Zwartbol}, and more recently, COVID-19 \cite{Lersy2021,Lin2020}. CMB size can range from 2 mm to 5-10 mm in diameter \cite{Greenberg2009}. These haemosiderin deposits are superparamagnetic, i.e., they have high magnetic susceptibility. Therefore, they produce inhomogeneity in the magnetic field of the surrounding area, making the decay of the local magnetic resonance imaging (MRI) signal faster. Hence, CMBs are typically seen as hypointense foci in MRI sequences sensitive to susceptibility effects such as gradient-recalled echo (GRE) T2$^*$-weighted imaging (T2$^*$WI) or susceptibility-weighted imaging (SWI). Compared to GRE T2$^*$WI, SWI increases the CMB detection rate from 20\% to 40\% \cite{Haller2018}.

Deep learning methods applied to CMB segmentation \cite{Liu2019,Chen2019,Dou2016,Al-masni2020} mostly consist of two-stage pipelines to search for potential positive objects and to differentiate true positives (TP) from false positives (FP). The state-of-the-art methods in \cite{Liu2019} uses a two-stage approach based on 3D-FRST \cite{Loy} and 3D-ResNet \cite{He_2016_CVPR} trained on high-resolution images and phases to achieve 95\% sensitivity, 71\% precision, and an average FP of 1.6 per scan. However, the accuracies drop to 73\% sensitivity, 67\% precision, and an average FP of 1.9 per scan, when applied to low-resolution scans \cite{Al-masni2020}. Due to learning from only 1-2 different datasets, these methods usually fail to generalize to data with other characteristics obtained in uncontrolled clinical settings.

Data augmentation is a common technique for addressing some of the domain shift issues in medical imaging \cite{ghazi2022fast}, but it is often not enough to cope with all differences in demographics, comorbidities, and acquisition parameters to achieve high generalization accuracy. Therefore, sometimes there is a need for including data with high variability in the training set, especially when the training data is small. However, this requires the acquisition and annotation of some MRI scans from different sites and parameters, which is a time-consuming and labor-intensive task and can cause data privacy challenges.

In this work, a robust deep-learning model is trained and applied to assess COVID-19 CMBs in a retrospective cohort study data, obtained from the capital region of Denmark \cite{Jimenez-Solem2021} and weakly labeled as case-control through matching. MultiResUNet architecture \cite{Ibtehaz2020} is trained on different heterogeneous datasets using the three orthogonal planes of each scan for CMB segmentation and detection. Two public datasets assigned for normal aging, stroke, and Alzheimer’s disease analysis \cite{Momeni2021,Dou2016} as well as an in-house dataset for COVID-19 assessment are used to train and evaluate the models. The obtained results show that the proposed method is robust to low-resolution images and achieves 78\% sensitivity and 80\% precision on the entire test set with an average FP of 1.6 per scan. To show the stability of the method on heterogeneous data, the pipeline is successfully applied to distinguish between the two weakly labeled groups.

\section{Method}
\label{sec:method}
Our study design comprises strategic data selection and preprocessing, data augmentation and training deep learning models, postprocessing for classifier fusion, and standard testing for the detection and group differentiation capability. The utilized deep learning framework includes MultiResUNet models \cite{Ibtehaz2020} trained individually on axial, sagittal, and coronal views, which are later fused for the 3-views agreement.
\iffalse
, as shown in \autoref{fig:schema}.

\begin{figure}[htb]
\centering
\includegraphics[scale=0.35]{figures/CMBpipeline.pdf}
\caption{The utilized CMB segmentation-detection framework.}
\label{fig:schema}
\end{figure}
% For the footnotes, use Times 9-point type, single-spaced.
\fi

\subsection{Data processing}
Two public datasets \cite{Momeni2021,Dou2016} and in-house data with SWI images are used in this study; the first data (DS1) is obtained from \cite{Momeni2021} and contains 57 scans from 30 patients with real CMBs (DS1r) and 3,700 scans from 118 patients with synthetic CMBs (DS1s); the second data (DS2) is obtained from \cite{Dou2016} and contains 20 scans from 20 subjects with real CMBs; the third dataset (DS3) is obtained from different hospitals of the capital region of Denmark and contains 40 scans from 40 patients with CMBs and 22 scans from 22 patients with no CMBs (DS3n). Moreover, 40 scans with no CMBs, matched with those 40 scans with CMBs, are used for testing the group distinguishability of the model.

The DS1 scans were acquired in a spatial resolution of 0.93 $\times$ 0.93 $\times$ 1.75 mm$^3$ using a 3T scanner with 20 ms echo time. The DS2 scans were obtained in a resolution of 0.45 $\times$ 0.45 $\times$ 1 mm$^3$ using a 3T scanner with 24 ms echo time. Finally, the DS3 scans were captured from different sites and scanners (1.5T and 3T) in various resolutions (from 0.2 $\times$ 0.2 $\times$ 1 mm$^3$ to 1 $\times$ 1 $\times$ 6 mm$^3$), to increase our model’s robustness and generalization accuracy.

Since the DS1 and DS2 only provide the CMB center positions on the SWI images, to train a multiview deep network, we need to create volumetric masks for the CMBs. We select patches with an isotropic voxel size of 10 mm$^3$ as the maximum CMB size according to \cite{Greenberg2009} around the CMB centers and estimate the fraction of microbleeds for each pixel using
\begin{equation*}
\alpha_{\text{CMB}} = \frac{I_{pixel} - I_{mean}}{ I_{center} - I_{mean}},
\label{meth_data_pve}
\end{equation*}
where $I_{pixel}$ is the current pixel’s intensity, $I_{center}$ is the center pixel’s intensity, and $I_{mean}$ is the mean intensity of the surrounding pixels in a 2 mm neighborhood. Pixels with $\alpha>0.65$ and $\alpha>0.52$ were labeled as CMBs in DS1 and DS2, respectively. The different $\alpha$ values were obtained due to the difference in echo time or tissue contrast.

The DS3 scans were annotated by clinical experts and weakly labeled by probabilities estimated using the electronic healthcare records \cite{Jimenez-Solem2021} based on logistic regression and odds ratio according to the UK Biobank study \cite{Lu2021a}. For testing group differentiation capability, we selected 40 scans with $p_{\text{CMB}} > 0.3$, annotated them, and matched them on 40 unannotated scans with $p_{\text{CMB}} < 0.01$ based on the magnetic field strength, echo time, slice thickness, and scanner model.

\subsection{CMB segmentation and detection}
A 3D framework is built based on 2D convolutional neural networks (CNNs) to segment and count the number of CMBs. The pipeline consists of MultiResUNet models \cite{Ibtehaz2020}, trained individually on three orthogonal views of the MRI scans and fused later for the mask volume reconstruction. The volumes are first reshaped to 256 $\times$ 256 $\times$ 256 with an isotropic voxel size of 1 mm$^3$. To benefit from the 3D information, thickened 2D slices are then formed by three consecutive images of each MRI scan as inputs to the 3-channel CNNs.

Several MRI-specific transforms are randomly applied to the training images for data augmentation \cite{ghazi2022fast}, including elastic deformation, bias field, rotation, flipping, blurring, motion ghosting, Gibbs ringing, and additive-multiplicative noise. Moreover, intensity normalization and contrast adjustment are applied to the network inputs. Finally, the output probability maps of the central slices from different views are fused by multiplying the three maps to obtain the label masks based on a full agreement between the three predictions. Any quantitative measures like the counts, position, and size of the detected CMBs can then be computed using the output map.

\section{Experiments and results}
\label{sec:experiment}

\subsection{Experimental setup}
Scans from different subjects are randomly partitioned into 70\%, 10\%, and 20\% for training, validation, and testing. To balance different datasets for training, only 10\% of DS3n slices and 5\% of DS1s slices are used, optimized using a grid search method on the validation set.

The networks, pretrained on medical data for skin lesion segmentation \cite{Ibtehaz2020}, are finetuned for CMB segmentation using the Adam Optimizer with cross-entropy loss function and a weight decay of $10^{-6}$. The base learning rate is set to $5\times10^{-4}$ and decreased by 10\% every 10 epochs. The hyperparameter values are obtained using a grid search algorithm on the validation set.

\subsection{Results and discussion}
The test prediction results of our model on different datasets are summarized in \autoref{tab:results_obj_metrics1}. As can be seen, our framework achieves a Dice similarity score (DSC) of 0.79, sensitivity of 0.78, precision of 0.80, and an average FP of 1.64 per scan. Note that CMB segmentation and detection in the wild (DS3) is very challenging, as the data is acquired from uncontrolled clinical settings with different imaging protocols. The high variations in the image parameters, anatomies, and patient conditions can affect the model's accuracy and robustness. The proposed method, trained on a few annotated DS3 scans, has achieved comparable results in the DS3 test set.

\begin{table}[t]
\mysizea
\centering
\caption{CMB segmentation and detection results on different test sets using the proposed framework.}
\vspace{0.1cm}
\begin{tabular}{lcccccc}
\toprule
\textbf{Dataset} & \textbf{TP/scan} & \textbf{FP/scan} & \textbf{FN/scan} & \textbf{DSC} & \textbf{Sensitivity} & \textbf{Precision} \\
\midrule
DS1r & 3.58 & 2.00 & 1.08 & 0.70 & 0.77 & 0.64 \\
DS1s & 8.11 & 1.75 & 1.72 & 0.82 & 0.83 & 0.82 \\
DS2 & 1.00 & 0.00 & 1.00 & 0.67 & 0.50 & 1.00 \\
DS3 & 8.57 & 1.43 & 5.00 & 0.73 & 0.63 & 0.86 \\
DS3n & 0.00 & 0.00 & 0.00 & 1.00 & NA & NA \\
\midrule
All & 6.75 & 1.64 & 1.92 & 0.79 & 0.78 & 0.80 \\
\bottomrule
\end{tabular}
\label{tab:results_obj_metrics1}
\end{table}

The test results of the DS3 show lower sensitivity and higher precision. The higher FN rate can be due to the difference in CMB distribution per scan compared to the other datasets, where many scans could have a single CMB while many others have a large number of CMBs. The higher precision can also be due to the higher resolution of DS3 scans compared to the alternatives. In addition, as expected, no CMBs were detected on DS3n, indicating that our method can be applied to distinguish between healthy and diseased scans. Finally, the lower accuracy obtained on DS2 can be because it has the smallest dataset size in terms of the number of scans and CMBs per scan.

Due to the different nature of the used datasets in the literature, we classify them as low-resolution (LR) and high-resolution (HR) SWI-based studies for a fair comparison. The HR studies on CMB segmentation involve \cite{Al-masni2020,Liu2019,Dou2016} with scans of spacings 0.5 $\times$ 0.5 $\times$ 2 mm$^3$, 0.45 $\times$ 0.5 $\times$ 1.2 mm$^3$ to 0.54 $\times$ 1.07 $\times$ 2.65 mm$^3$, and 0.45 $\times$ 0.45 $\times$ 2 mm$^3$, respectively, while the LR one is obtained from \cite{Al-masni2020} with scans of spacing 0.8 $\times$ 0.8 $\times$ 2 mm$^3$. In contrast, our LR test set contains scans with spacings of 0.89 $\times$ 0.89 $\times$ 2 mm$^3$ (DS3) and 0.93 $\times$ 0.93 $\times$ 1.75 mm$^3$ (DS1), while the HR test set includes scans with spacings of 0.3 $\times$ 0.3 $\times$ 2 mm$^3$ to 0.34 $\times$ 0.34 $\times$ 4 mm$^3$ (DS3) and 0.45 $\times$ 0.45 $\times$ 1 mm$^3$ (DS2). It should also be noted that \cite{Al-masni2020,Liu2019} benefit from the phase images for segmentation. \autoref{tab:results_obj_metrics1} shows the prediction results of different models on various scan resolutions of the test sets. As seen, our model achieves better results on the LR subset while obtaining poor sensitivity on the HR test set. The reason for such behavior compared to the state-of-the-art could be because the scan resolutions in different studies do not match exactly, our model is trained on much more LR scans ($83\%$ LR scans vs. $17\%$ HR scans), and we do not use phase images for CMB segmentation and detection. Moreover, the HR dataset of DS2 used in this study only includes a small publicly available subset of the one used in \cite{Dou2016}.

\begin{figure}[t]
	\centering
	% include second image
	\includegraphics[width=1.0\linewidth,height=0.42\linewidth]{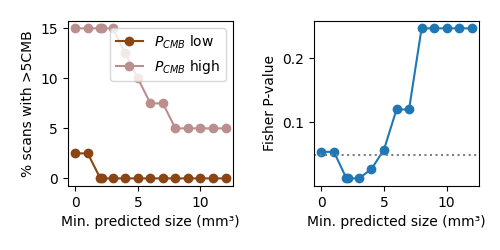}  
	\caption{The CMB frequency of the two groups and their statistical significance for the filtered CMB size. As seen, the two groups with more than 5 CMB are significantly different for the CMB size smaller than 5 mm$^3$.}
	\label{fig:results_varying_size_cph}
\end{figure}

Finally, to see how the weakly labeled groups of our COVID-19 study are statistically significantly different in terms of the number of detected CMBs, we use two unannotated scan subsets labeled as non-CMB ($p_{\text{CMB}} < 0.01$) and CMB ($p_{\text{CMB}} > 0.3$). On average per scan, we found 0.62 CMB in the non-CMB scans and 2.60 CMB in the CMB scans, which were statistically significantly different with a $p = 0.001$ using the Wilcoxon signed-rank test. Nest, we set clinical criteria for illness based on 5 CMBs per scan and attempt to distinguish between the groups with more or less than 5 CMBs. To minimize the FP effects, we filter the predicted CMBs by the minimum clinical size of 4.2 mm$^3$ with a 2 mm diameter. \autoref{fig:results_varying_size_cph} shows the difference between the two groups for the filtered CMB size, where the CMB frequency of the two filtered groups is statistically significantly different with a $p = 0.013$ using Fisher’s exact test.

\begin{table}[t]
\mysizea
\centering
\caption{Test CMB segmentation and detection results of different state-of-the-art models on various scan resolutions.}
\vspace{0.1cm}
\begin{tabular}{lccccc}
\toprule
\textbf{Model} & \textbf{Type} & \textbf{Phase} & \textbf{FP/scan} & \textbf{Sensitivity} & \textbf{Precision} \\
\midrule
YOLO + 3D-CNN \cite{Al-masni2020} & LR & Yes & 1.89 & 0.78 & 0.67  \\
& & & & & \\
YOLO + 3D-CNN \cite{Al-masni2020} & HR & Yes & 1.42 & 0.88 & 0.62  \\
& & & & & \\
3D-FRST + 3D-ResNet \cite{Liu2019} & HR & Yes & 1.6 & 0.95 & 0.71  \\
& & & & & \\
3D-FCN + 3D-CNN \cite{Dou2016} & HR & No & 2.74 & 0.91 & 0.44  \\

\midrule
The proposed & LR & No & 1.92 & 0.80 & 0.80  \\
The proposed & HR & No & 0.30 & 0.49 & 0.87 \\
\bottomrule
\end{tabular}
\label{tab:results_obj_metrics2}
\end{table}

\section{Conclusion}
\label{sec:conclusion}
In this work, we developed a deep learning-based framework based on MultiResUNet architecture using different heterogeneous datasets and MRI-specific data augmentation for CMB segmentation and detection and assessed COVID-19 CMBs in a retrospective cohort study. The proposed pipeline was tested on uncontrolled clinical data and outperformed the state-of-the-art methods, especially when applied to low-resolution images. Moreover, the proposed pipeline was able to distinguish between two weakly labeled groups of the heterogeneous COVID-19 data, and the results were statistically significantly different in terms of the number of detected CMBs.

\section{Acknowledgments}
\label{sec:acknowledgments}

This project has received funding from VELUX FONDEN and Innovation Fund Denmark under grant numbers 9084-00018B and 1063-00014B, and Pioneer Centre for AI, Danish National Research Foundation, grant number P1.
This study was conducted using human subject data. Ethical approval was not required for the open-access data as confirmed by the license attached to the data. The Danish cohort study was approved by the relevant legal and ethics boards, including the Danish Patient Safety Authority (Styrelsen for Patientsikkerhed, approval $\#$31-1521-257) and the Danish Data Protection Agency (Datatilsynet, approval $\#$P-2020-320).

\bibliographystyle{IEEEbib-abbrev}
\bibliography{refs}

\end{document}